%% file: SocInfo_2017_Effects_arXiv.tex
\pdfoutput=1

\documentclass{llncs}

\usepackage{amsmath}
\usepackage{graphicx}
\usepackage{subfig}
\usepackage[margin=false,inline=true]{fixme}
\usepackage{multirow}
\usepackage{array}
\newcolumntype{C}[1]{>{\centering\arraybackslash}m{#1}}

\usepackage{hyperref}

\newcommand{\ER}{Erd\H{o}s-R\'{e}nyi }
\renewcommand{\b}{\mathbf}
\renewcommand{\vec}[1]{\boldsymbol{\mathbf{#1}}}

\begin{document}

\title{Effects of Contact Network Models on Stochastic Epidemic Simulations}

\titlerunning{Effects of Network Models on Epidemics}  

\author{Rehan Ahmad \and Kevin S.~Xu}

\institute{EECS Department, University of Toledo, Toledo, OH  43606, USA\\
\email{Rehan.Ahmad@rockets.utoledo.edu}, \email{Kevin.Xu@utoledo.edu}}

\maketitle              

\begin{abstract}
The importance of modeling the spread of epidemics through a population has led to the development of mathematical models for infectious disease propagation. 
A number of empirical studies have collected and analyzed data on contacts between individuals using a variety of sensors. Typically one uses such data to fit a  probabilistic model of network contacts over which a disease may propagate. 
In this paper, we investigate the effects of different contact network models with varying levels of complexity on the outcomes of simulated epidemics using a stochastic Susceptible-Infectious-Recovered (SIR) model. 
We evaluate these network models on six datasets of contacts between people in a variety of settings. 
Our results demonstrate that the choice of network model can have a significant effect on how closely the outcomes of an epidemic simulation on a simulated network match the outcomes on the actual network constructed from the sensor data. 
In particular, preserving degrees of nodes appears to be much more important than preserving cluster structure for accurate epidemic simulations.
\end{abstract}

\input{intro.tex}

\input{related.tex}

\input{datasets.tex}

\input{methods.tex}

\input{results.tex}

\input{discussion.tex}

\bibliographystyle{splncs03}
\bibliography{SocInfo_2017_Effects}

\end{document}

%% file: intro.tex
\section{Introduction}

The study of transmission dynamics of infectious diseases often involves simulations using stochastic epidemic models.  
In a compartmental stochastic epidemic model, transitions between compartments occur randomly with specified probabilities. 
For example, in a stochastic Susceptible-Infectious-Recovered (SIR) model \cite{Britton2010,Greenwood2009}, a person may transition from S to I with a certain probability upon contact with an infectious person, or a person may transition from I to R with a certain probability to simulate recovering from the disease.

The reason for the spread of infection is contact with the infectious individual. 
Hence, the contact network in a population is a major factor in the transmission dynamics. 
Collecting an actual contact network over a large population is difficult because of limitations in capturing all the contact information. 
This makes it necessary to represent the network with some level of abstraction, e.g.~using a statistical model. 
A variety of statistical models for networks have been proposed \cite{Goldenberg:2010:SSN:1734794.1734795}; such models can be used to simulate contact networks that resemble actual contact networks.

Our aim in this paper is to evaluate different models for contact networks in order to find the best model to use to simulate contact networks that are close to an actual observed network. 
We do this by comparing the disease dynamics of a stochastic SIR model over the simulated networks with the disease dynamics over the actual network. 
One commonly used approach is to compare the epidemic size at the end of the simulation, i.e.~what fraction of the population caught the disease \cite{Machens2013,StehlÃ©2011}. 
A drawback of this approach is that it only considers the steady-state outcome and not the dynamics of the disease as it is spreading.

\begin{figure}[t]
\centering
\subfloat[Susceptible]{\includegraphics[width=1.5in]{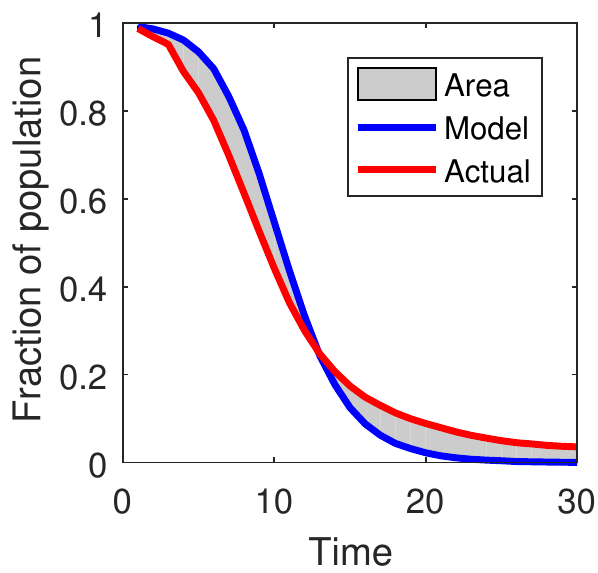}}
\quad
\subfloat[Infectious]{\includegraphics[width=1.5in]{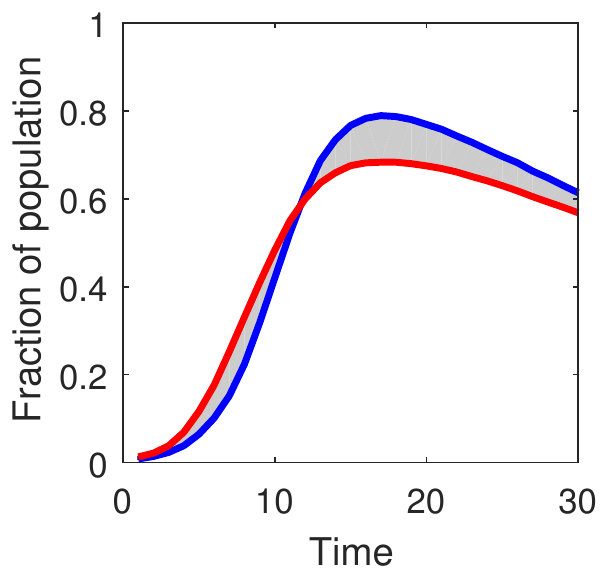}}
\quad
\subfloat[Recovered]{\includegraphics[width=1.5in]{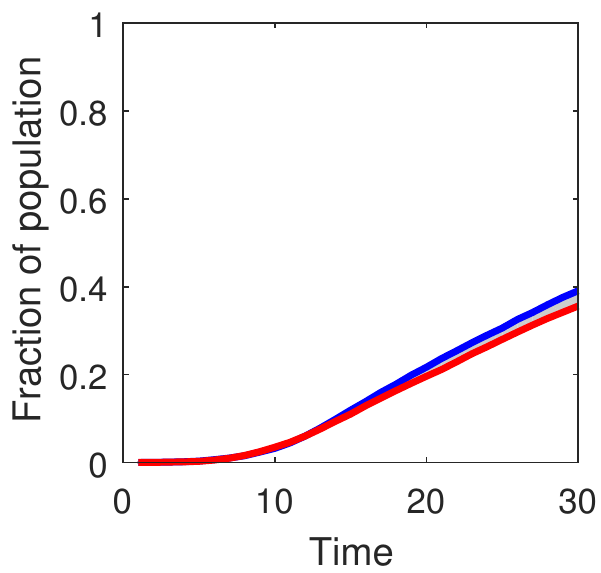}}
\caption{For each of the susceptible (S), infectious (I), and recovered (R) compartments, the mean curve for simulations on the model (shown in blue) is compared to the mean curve for simulations on the actual network (shown in red). The closeness between the model and actual network is given by the sum of the shaded areas between the curves for each compartment (smaller is better).}
\label{fig:AreaSIR}
\end{figure}

We propose to compare the dynamics at each time instant in the simulation by calculating the area between the mean SIR curves for the epidemic over the simulated and actual networks, shown in Fig.~\ref{fig:AreaSIR}. 
A small area indicates that the dynamics of the epidemic over the simulated contact networks are close to those of the actual network. We use this approach to compare four contact network models (in increasing order of number of parameters): the \ER model, the degree model, the stochastic block model, and the degree-corrected stochastic block model. Our experiment results over six different real network datasets suggest that the degree-corrected stochastic block model provides the closest approximation to the dynamics of an epidemic on the actual contact networks. Additionally, we find that preserving node degrees appears to be more important than preserving community structure for accuracy of epidemic simulations.

%% file: related.tex
\section{Related Work}

A significant amount of previous work deals with the duration \cite{Smieszek2009a}, frequency \cite{larson07}, and type \cite{eames08,Smieszek2009} of contacts in a contact network. 
These findings are often incorporated into simulations of epidemics over different types of contact models.
The R package EpiModel \cite{jenness2017package} allows for simulation of a variety of epidemics over temporal exponential random graph models for contact networks and has been used in studies of various different infectious diseases including HIV \cite{doi:10.1093/infdis/jiw223}.

There has also been prior work simulating the spread of disease over a variety of contact network models with the goal of finding a good approximation to the actual high resolution data in terms of the epidemic size, i.e.~the final number of people infected \cite{Machens2013,StehlÃ©2011}. 
Such work differs from our proposed area metric, which considers the dynamics as the disease is spreading and not just the steady-state outcome. 
In \cite{Bioglio2016}, the authors use the squared differences between the I curves (fraction of infectious individuals) of an epidemic model on simulated contact networks and on an actual contact network to calibrate parameters of the epidemic model when used on simulated contact networks. 
Although this metric does consider the dynamics of the epidemic, our proposed metric also involves the S and R curves for a more complete evaluation of population dynamics.

%% file: datasets.tex
\section{Datasets}

\begin{table}[t]
\centering
\caption{Summary statistics from datasets used in this study.}
\label{tab:DataSets}

\begin{tabular}{ccC{0.6in}C{0.5in}cC{0.55in}c}
\hline
                       & HYCCUPS & Friends \& Family & High School  & Infectious   & Primary School & HOPE  \\
\hline
Number of nodes        & $43$    & $123$             & $126$        & $201$        & $242$          & $1178$    \\
Sensor type            & Wi-Fi   & Bluetooth         & RFID         & RFID         & RFID           & RFID      \\
Proximity range        & N/A     & $5$ m             & $1$--$1.5$ m & $1$--$1.5$ m & $1$--$1.5$ m   & Room \\
Graph density          & $0.326$ & $0.228$           & $0.217$      & $0.0328$     & $0.285$        & $0.569$   \\
Clustering coefficient & $0.604$ & $0.496$           & $0.522$      & $0.459$      & $0.480$        & $0.748$   \\
Average degree         & $14.0$  & $27.8$            & $27.1$       & $6.56$       & $68.7$         & $671$     \\
Maximum degree         & $28$	 & $73$              & $55$	        & $21$	       & $134$          & $1072$    \\
\hline
\end{tabular}
\end{table}

We consider a variety of contact network datasets in this paper. 
Table \ref{tab:DataSets} shows summary statistics for each dataset along with the sensor type. The HYCCUPS dataset was collected at the University Politehnica of Bucharest in 2012 using a background application for Android smartphones that captures a device's encounters with Wi-Fi access points \cite{Marin2012}. 
The Friends \& Family (F\&F) dataset was collected from the members of a residential community nearby a major research university using Android phones loaded with an app that records many features including proximity to other Bluetooth devices \cite{Aharony2011643}.
The High School (HS) dataset was collected among students from $3$ classes in a high school in Marseilles, France \cite{10.1371/journal.pone.0107878} using wearable sensors that capture face-to-face proximity for more than $20$ seconds.
The Infectious dataset was collected at a science gallery in Dublin using wearable electronic badges to sense sustained face-to-face proximity between visitors. \cite{Isella:2011qo}. 
We use data for one arbitrarily selected day (April 30) on which $201$ people came to visit. 
The Primary School (PS) dataset was collected over $232$ students and $10$ teachers at a primary school in Lyon, France in a similar manner to the HS dataset \cite{Gemmetto2014}. 
Lastly, the HOPE dataset is collected from the Attendee Meta-Data project at the seventh Hackers on Planet Earth (HOPE) conference \cite{hope-amd-20080807}. 
We create a contact network where the attendees at each talk form a clique; that is, each person is assumed to be in contact with every other person in the same room, hence why this network is much denser.

%% file: methods.tex
\section{Methods}
We construct actual networks from the datasets by connecting the individuals (nodes) with an edge if they have a contact at any point of time. We evaluate the quality of a contact network model for simulations of epidemics by conducting the following steps for each dataset:
\begin{enumerate}
\item Simulate $5,000$ epidemics over the actual network.
\item Fit contact network model to actual network.
\item Simulate $100$ networks from contact network model. For each simulated network, simulate $50$ epidemics over the network for $5,000$ epidemics total.
\item Compare the results of the epidemic simulations over the actual network with those over the simulated networks.
\end{enumerate}

These steps are repeated for each contact network model that we consider. We describe the stochastic epidemic model we use to simulate epidemics in Section \ref{sec:EpiModel} and the contact network models we use in Section \ref{sec:NetModel}.
To get a fair evaluation of the dynamics of epidemics spreading over different contact network models, all of the parameters which are not related to the contact network model, e.g.~probability of infection and probability of recovery are kept constant. Our aim is to single out the effect of using a particular contact network model while simulating an epidemic.

\subsection{Stochastic Epidemic Model}
\label{sec:EpiModel}
An actual infection spread in a population experiences randomness in several factors which may aggravate or inhibit the spread. This is considered in stochastic epidemic models. The initial condition is, in general, to have a set of infectious individuals, while the rest of the population is considered susceptible. 
We consider a discrete-time process, where at each time step, the infectious individuals can spread the disease with some probability of infection to susceptible individuals they have been in contact with. Also, the infectious individuals can recover from the disease with some probability independent of the individuals' contacts with others. This model is known as the stochastic SIR model and is one of the standard models used in epidemiology \cite{Britton2010,Greenwood2009}. 

We randomly choose $1$ infectious individual from the population as the initial condition and simulate the epidemic over $30$ time steps.
We set the probability of infection for every interaction between people to be $0.025$. The probability of recovery is also set to be $0.025$. Note that the rate at which the disease spreads across the population is dependent not only on the infection probability but also the topology of the contact network; thus, by fixing these probabilities, we are exploring only the effects of the contact network.

\subsection{Contact Network Models}
\label{sec:NetModel}
In practice, it is extremely difficult to obtain accurate contact network data. An alternative is to simulate a contact network by using a statistical network model. We consider several such models, which we briefly describe in the following. 
We refer interested readers to the survey by Goldenberg et al.~\cite{Goldenberg:2010:SSN:1734794.1734795} for details.

\subsubsection{\ER (E-R) Model}

In the E-R model, an edge between any two nodes is formed with probability $p$ independent of all other edges. To fit the E-R model to a network, set the single parameter, the estimated edge probability $\hat{p} = M/\binom{N}{2}$, where $N$ and $M$ denote the number of nodes and edges in the actual network, respectively. By doing so, the expected number of edges in the E-R model will be $\binom{N}{2}\hat{p} = M$, the number of edges in the actual network.

\subsubsection{Degree Model}
In several network models, including the configuration model and preferential attachment models, the edge probability depends upon the degrees of the nodes it connects \cite{Newman:2010:NI:1809753}. We consider a model that preserves the expected rather than actual degree of each node, often referred to as the Chung-Lu model \cite{Chung2002}. In this model, the probability of an edge between two nodes is proportional to the product of their node degrees, and all edges are formed independently. 
The model has $N$ parameters, the expected degrees of each node.

To fit the degree model to a network, we compute the degrees of all nodes to obtain the degree vector $\vec{d}$. We then set the estimated edge probabilities $\hat{p}_{ij} = \alpha d_i d_j$, where the constant $\alpha$ is chosen so that the sum of all edge probabilities (number of expected edges) is equal to the number of edges in the actual network.

\subsubsection{Stochastic Block Model (SBM)} 
In the SBM \cite{Holland1983}, the network is divided into disjoint sets of individuals forming $K$ communities. The probability of edge formation between two nodes depends only upon the communities to which they belong. This model takes as input a vector of community assignments $\vec{c}$ (length $N$) and a matrix of edge formation probabilities $\Phi$ (size $K \times K$), where $\phi_{ab}$ denotes the probability that a node in community $a$ forms an edge with a node in community $b$, independent of all other edges. 
For an undirected graph, $\Phi$ is symmetric so the SBM has $N + \binom{K+1}{2}$ parameters in total.

To estimate community assignments, we use a regularized spectral clustering algorithm \cite{Qin2013} that is asymptotically consistent and has been demonstrated to be very accurate in practice. 
We select the number of communities using the eigengap heuristic \cite{VonLuxburg2007}. 
Once the community assignments $\hat{\vec{c}}$ are estimated, the edge probabilities can be estimated by $\hat{\phi}_{ab} = m_{ab} / n_{ab}$, where $m_{ab}$ denotes the number of edges in the block formed by the communities $a,b$ in the observed network, and $n_{ab}$ denotes the number of possible edges in the block \cite{PhysRevE.83.016107}. 

\subsubsection{Degree-corrected Stochastic Block Model (DC-SBM)}
The DC-SBM is an extension to the SBM in a way that incorporates the concepts of the degree model within an SBM \cite{PhysRevE.83.016107}. The parameters of the DC-SBM are the vector of community assignments $\vec{c}$ (length $N$), a node-level parameter vector $\vec{\theta}$ (length $N$), and a block-level parameter matrix $\Omega$ (size $K \times K$). In a DC-SBM, an edge between a node $i \in a$ (meaning node $i$ is in community $a$) and node $j \in b$ is formed with probability $\theta_i \theta_j \omega_{ab}$ independent of all other edges. 
$\Omega$ is symmetric, so the DC-SBM has $2N + \binom{K+1}{2}$ parameters in total.

To fit the DC-SBM to an actual network, we first estimate the community assignments in the same manner as in the SBM using regularized spectral clustering. 
We then estimate the remaining parameters to be $\hat{\theta}_i = d_i / \sum_{j \in a} d_j$, for node $i \in a$, and $\hat{\omega}_{ab} = m_{ab}$ \cite{PhysRevE.83.016107}.
Using these estimates, we arrive at the estimated edge probabilities $\hat{p}_{ij} = \hat{\theta}_i \hat{\theta}_j \hat{\omega}_{ab}$. 

%% file: results.tex
\section{Results}

To evaluate the quality of a contact network model, we compare the mean SIR curves resulting from epidemic simulations on networks generated from that model to the mean SIR curves from epidemic simulations on the actual network. 
If the two curves are close, then the network model is providing an accurate representation of what is likely to happen on the actual network. 

To measure the closeness of the two sets of mean SIR curves, we use the sum of the areas between each set of curves as shown in Fig.~\ref{fig:AreaSIR}. By measuring the area between the curves rather than just the final outcome of the epidemic simulation (e.g.~the fraction of recovered people after the disease dies out as in \cite{Machens2013,StehlÃ©2011}), we capture the difference in transient dynamics (e.g.~the rate at which the infection spreads) rather than just the difference in final outcomes.

\begin{figure}[t]
\centering
\subfloat[]{\includegraphics[width=2.25in]{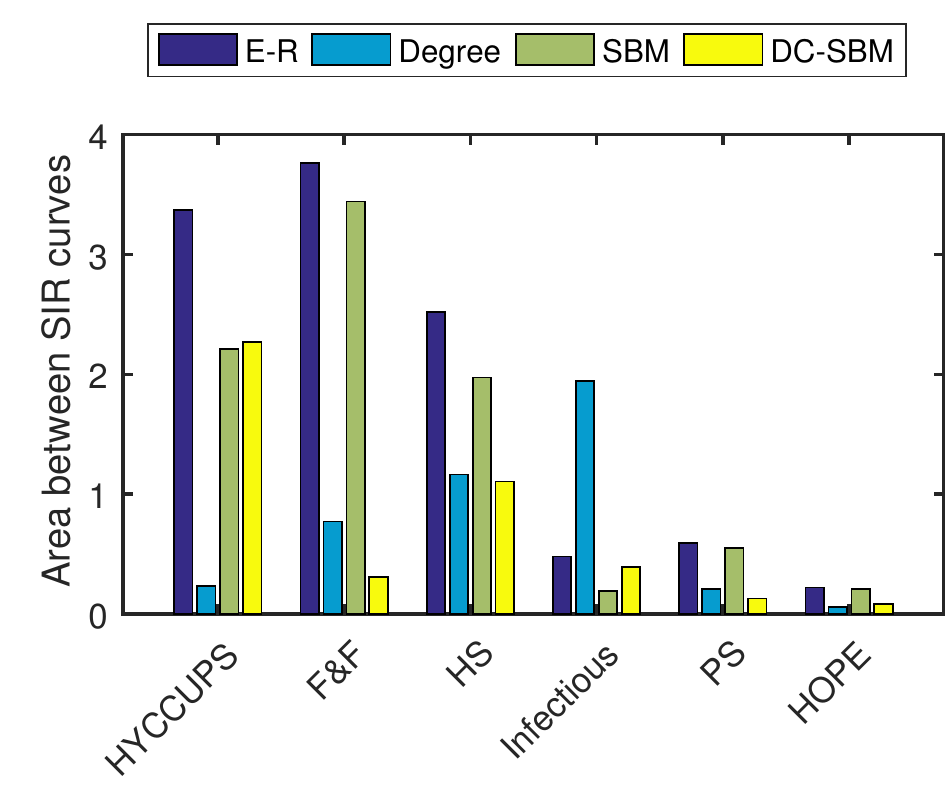}
 \label{fig:AreaBars3}}
\quad
\subfloat[]{\includegraphics[width=2.25in]{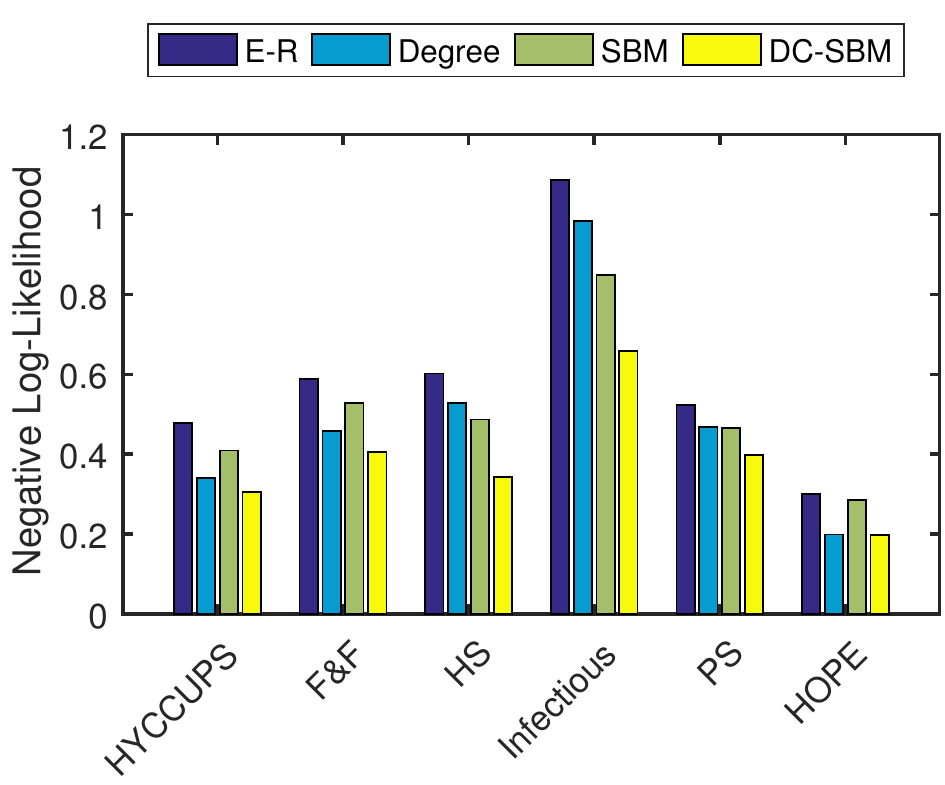}
 \label{fig:LoglikBars3}}
\caption[]{Comparison of \subref{fig:AreaBars3} area between SIR curves of each model with respect to actual network for each dataset and \subref{fig:LoglikBars3} negative log-likelihood per node pair for each model (lower is better for both measures). The DC-SBM model appears to be the best model according to both quality measures, but the two measures disagree on the quality of the degree model compared to the SBM.}
\end{figure}

The area between the SIR curves for each model over each dataset is shown in Fig.~\ref{fig:AreaBars3}.
According to this quality measure, the DC-SBM is the most accurate model on F\&F, HS, and PS; the degree model is the most accurate on HYCCUPS and HOPE; and the SBM is most accurate on Infectious. 
However, the SBM appears to be only slightly more accurate than the E-R model overall, despite having $N+\binom{K+1}{2}$ parameters compared to the single parameter E-R model. 
The contact network models were most accurate on the HOPE network, which is the 
densest, causing the epidemics to spread rapidly. 

We compute also the log-likelihood for each contact network model on each dataset, shown in Fig.~\ref{fig:LoglikBars3}. 
To normalize across the different sized networks, we compute the log-likelihood per node pair. 
Since all of the log-likelihoods are less than $0$, we show the negative log-likelihood (i.e.~lower is better) in Fig.~\ref{fig:LoglikBars3}. 
Unsurprisingly, the DC-SBM, with the most parameters, also has the highest log-likelihood, whereas the relative ordering of the log-likelihoods of the degree model and SBM, both with roughly the same number of parameters, vary depending on the dataset.

\begin{table}[t]
\centering
\caption{Quality measures (lower is better) averaged over all datasets for each model. Best model according to each measure is shown in bold.}
\label{tab:AvgMetrics}
\setlength\tabcolsep{0.5em}
\begin{tabular}[b]{ccccc}
\hline
Quality Measure                       & E-R      & Degree   & SBM      & DC-SBM       \\
\hline
Area between SIR curves               & $1.82$   & $0.73$   & $1.43$   & $\b{0.71}$   \\
Negative log-likelihood per node pair & $0.597$  & $0.496$  & $0.504$  & $\b{0.385}$  \\
Number of parameters                  & $\b{1}$  & $319$    & $328$    & $647$        \\
\hline
\end{tabular}
\end{table}  

Both the proposed area between SIR curves and the log-likelihood can be viewed as quality measures for a contact network model. 
A third quality measure is given by the number of parameters, which denotes the simplicity of the model. 
A simpler model is generally more desirable to avoid overfitting. 
These three quality measures for each model (averaged over all datasets) are shown in Table \ref{tab:AvgMetrics}. 
The DC-SBM achieves the highest quality according to the area between SIR curves and the log-likelihood at the expense of having the most parameters. 
On the other hand, the E-R model has only a single parameter but is the worst in the other two quality metrics. 
Interestingly, the degree model and SBM appear to be roughly equal in terms of the number of parameters and log-likelihood, but the area between SIR curves for the two models differs significantly. 
This suggests that the degree model may be better than the SBM at reproducing features of contact networks that are relevant to disease propagation. 

%% file: discussion.tex
\section{Discussion}
\label{sec:Discussion}

The purpose of our study was to evaluate the effects of contact network models on the results of simulated epidemics over the contact network. 
While it is well-known and expected that more complex models for contact network topology do a better job of reproducing features of the contact network such as degree distribution and community structure, we demonstrated that, in general, they also result in more accurate epidemic simulations. 
That is, the results of simulating an epidemic on a more complex network model are usually closer to the results obtained when simulating the epidemic on the actual network than if we had used a simpler network model. 
Moreover, models that preserve node degrees are shown to produce the most accurate epidemic simulations. Unlike most prior studies such as \cite{Machens2013,StehlÃ©2011}, we measure the quality of a network model by its area between SIR curves compared to the SIR curve of the actual network, which allows us to capture differences while the disease is still spreading rather than just the difference in the final outcome, i.e.~how many people were infected.

Our findings suggest that the degree-corrected stochastic block model (DC-SBM) is the best choice of contact network model in epidemic simulations because it resulted in the minimum average area between SIR curves. 
Interestingly, using the degree model resulted in an average area between SIR curves to be only slightly larger than the DC-SBM despite having less than half as many parameters, as shown in Table \ref{tab:AvgMetrics}. 
 The SBM (without degree correction) also has half as many parameters as the DC-SBM, but has over twice the area between SIR curves. 
We note that the difference between the degree model and the SBM \emph{cannot} be observed using log-likelihood as the quality measure, as both models are very close in log-likelihood. 
This leads us to believe that preserving degree has a greater effect on accuracy of epidemic simulations than preserving community structure. 
Furthermore, this finding demonstrates that one cannot simply evaluate the accuracy of a contact network model for epidemic simulations only by examining goodness-of-fit on the actual contact network! 

In practice, one cannot often collect high-resolution contact data on a large scale, so having accurate contact network models is crucial to provide realistic network topologies on which we can simulate epidemics. 
In this paper, we estimated the parameters for each contact network model using the contact network itself, which we cannot do in practice because the contact network is often unknown. 
As a result, one would have to estimate the model parameters from prior knowledge or partial observation of the contact network, which introduces additional error that was not studied in this paper. 
It would be of great interest to perform this type of sensitivity analysis to identify whether the DC-SBM and degree model are still superior even when presented with less accurate parameter estimates. 
Also, there is a risk of overfitting in more complex models which should be examined in a future extension of this work. 
Both issues could potentially be addressed by considering hierarchical Bayesian variants of network models such as the degree-generated block model \cite{Zhu2012}, which add an additional generative layer to the model with a smaller set of hyperparameters. 

Another limitation of this study is our consideration of static unweighted networks. 
Prior work \cite{karimi2013threshold,Machens2013,Smieszek2009a,StehlÃ©2011} has shown that it is important to consider the time duration of contacts between people, which can be reflected as weights in the contact network, as well as the times themselves, which can be accommodated by using models of dynamic rather than static networks, such as dynamic SBMs \cite{Xu2014a}. 
We plan to expand this work in the future by incorporating models of weighted and dynamic networks to provide a more thorough investigation.